\documentclass[doublecol]{epl2} 
\usepackage{amsfonts}
\usepackage{graphicx}
\usepackage{amsmath}
\usepackage{cuted}
\usepackage{comment}
\usepackage{mathtools}

\begin{document}
\title{Gauge theory applied to magnetic lattices}

\shorttitle{Gauge theory applied to magnetic lattices} 

\author{A. Di Pietro\inst{1,2} \and P. Ansalone\inst{1} \and V. Basso \inst{1} \and A. Magni \inst{1} \and G. Durin \inst{1}}
\shortauthor{A. Di Pietro \etal}

\institute{                    
  \inst{1} INRIM - Strada delle Cacce 91, 10135 Torino, Italy\\
  \inst{2} Politecnico di Torino - Corso Duca degli Abruzzi, 24, 10129 Torino, Italy 
}
\pacs{nn.mm.xx}{First pacs description}
\pacs{nn.mm.xx}{Second pacs description}
\pacs{nn.mm.xx}{Third pacs description}

\abstract{
Micromagnetic exchange is usually derived by performing the continuum limit of the Heisenberg model on a cubic lattice, where the exchange integrals are assumed to be identical for all nearest neighbors. This limitation normally imposes the use of a microscopic theory to explain the appearance of higher order magnetic interactions such as the Dzyaloshinskii-Moriya interaction (DMI). In this paper we combine graph- and gauge field- theory to simultaneously account for the symmetries of the crystal, the effect of spin-orbit coupling and their interplay on a micromagnetic level.  We obtain a micromagnetic theory accounting for the crystal symmetry constraints at all orders in exchange and show how to successfully predict the
form of micromagnetic DMI in all 32 point groups.
}

\maketitle

\section{Introduction}
The Heisenberg model \cite{STA-68, Nowak2007, 10.1088/978-0-7503-1074-1ch2} is a low energy limit of the more general Hubbard model \cite{Cleveland1976,Hoffmann2020} and it allows to translate the complexity of quantum mechanical exchange in a geometrical framework where the degrees of freedom are localized magnetic moments. Understanding the Heisenberg model beyond ordinary ferromagnetic/antiferromagnetc exchange is of central scientific relevance as higher order interactions are at the core of most of the exotic physical phenomena that could be harnessed in future spintronics devices \cite{Vedmedenko2020,Luo2021,Walker2021}. As an example, the Dzyaloshinskii-Moriya interaction (DMI) \cite{MOR-60,DZYALOSHINSKY1958241} is known to stabilize skyrmionic/antiskyrmionic structures \cite{Fert2013,Nagaosa2013} which hold promise as information carriers in novel spintronic memory devices. At the same time, micromagnetic solvers \cite{Vansteenkiste2014} that can then be employed to simulate the magnetization dynamics on the micro-scale rely on continuum formulations making the generalization of the Heisenberg model and it’s higher order extensions to the continuum \cite{Brown1963} of central importance. While the inclusion of higher order interactions in the Heisenberg model can be done by considering the low energy limit of increasingly complicated multi-band Hubbard models \cite{Hoffmann2020}, a general approach to derive higher order interactions in a micromagnetic framework is still missing. Micromagnetic exchange is normally derived starting from the extended Heisenberg model on a thin film
\begin{equation}
    \mathcal{H} = - \sum_{<i,j>} J_{ij} \bm{s}_i \cdot \bm{s}_j + \bm{D}_{ij} \cdot (\bm{s}_i \times \bm{s}_j) . \label{Heisenberg}
\end{equation}
where $\bm{s}_i$ represents a magnetic moment located at lattice site $i$, $J_{ij}$ represents the exchange integral between magnetic moments $i,j$ and $\bm{D}_{ij}$ is the DMI contribution to the Heisenberg Hamiltonian. 
The ordinary continuum limit of this Hamiltonian is performed separately on the symmetric part \cite{Kronmuller2019} $J_{ij}( \bm{s}_i \cdot \bm{s}_j)$ and the anti-symmetric part \cite{Thiaville2012} $\bm{D}_{ij} \cdot (\bm{s}_i \times \bm{s}_j) $. This yields the micromagnetic exchange energy functional
\begin{align}
     &E_{ex}[\mathbf{m} , \nabla \mathbf{m}] = \int_{\Omega_V} \big\{  A\mid \nabla \mathbf{m}\mid^2 \nonumber \\ &- 2 \bm{D}\cdot [ \bm{m}(\nabla \cdot \bm{m}) - (\nabla \cdot \bm{m})\bm{m}] \big\} \,\upd^3 \bm{r}
\end{align}
where $\Omega_V$ is the volume of the magnetic sample, $A$ represents the exchange stiffness of the material and $\bm{D}$ is the micromagnetic DMI vector. 
The generalization of DMI \cite{Rossler2011,Bogdanov2002} to arbitrary lattices is done phenomenologically and neglects the connection between ordinary symmetric exchange and higher order interactions. The generalized expression one finds in the literature is \cite{Hoffmann2020,Ullah2019} 
\begin{align}
     E_{ex}[\mathbf{m} , \nabla \mathbf{m}] = \int_{\Omega_V} \big\{  A\mid \nabla \mathbf{m}\mid^2 +  \hat{\bm{Q}}\mathcal{M}(\bm{m})\big\}\,\upd^3 \bm{r} 
\end{align}
 where $\hat{\bm{Q}}\mathcal{M}(\bm{m}) = \sum_{A,C}\hat{\bm{Q}}_{AC} \mathcal{M}_{AC}$ constitutes the DMI energy of the system \cite{Hoffmann} and is represented as the contraction of the DMI tensor and the chirality $\mathcal{M}(\bm{m}) = \nabla \bm{m} \times \bm{m}$ of the material. Performing the continuum limit in this way poses some issues though as one neglects the fact that higher order interactions are intimately related to the lower order ones as they come from the low energy limit of a more general energy functional. In this work we propose an alternative procedure to the continuum limit of the Heisenberg model \cite{Kronmuller2019} that employs graph theory to systematically account for lattices of arbitrary point group symmetry and local $SO(3)$ gauge invariance of the micromagnetic energy functional to account for the appearance of higher order interactions \cite{Ansalone2021,Basso2020a,Guslienko2016}. The outcome is going to be a continuum limit that naturally represents the exchange interaction energy in the most general form at all orders
 \begin{equation}
  E_{ex} = \displaystyle \int_{\Omega_V}  \big\{ \, \Xi_{AC}  \,\partial^A m^B  \partial^C m_B -  Q_{Al}\mathcal{M}^{lA} \, \big\} \, \upd^3 \bm{r}     \label{eq:main_res}
 \end{equation}
 where $\Xi_{AC}$ is the anisotropic symmetric exchange tensor and $Q_{Al}$ is the DMI tensor. 
The structure of the paper is as follows: in the first Section, we reformulate the continuum limit of exchange using graph theory and we rewrite exchange in a form that keeps track of the lattice beyond the simple cubic case. In the second Section we require local $SO(3)$ gauge invariance to account for the appearance of the DMI tensor \cite{Ansalone2022}. The Neumann principle of crystallography allows us to easily derive the non-vanishing components of the anisotropic exchange and micromagnetic DMI for all 32 crystallographic point groups. In the last Section we discuss our result and validate our predictions by comparing them with several experimental systems. In the Appendix we review the Taylor expansion on discrete lattices and some key concepts of graph theory. In the following we are going to employ the Einstein summation convention when the summation sign is not explicitly shown.

\section{Continuum limit of Heisenberg exchange on arbitrary lattices \label{Sec:Heis_arbitrary}} 
\begin{figure*}
    \centering
    \includegraphics[trim={0 3cm 0 0},clip , width=0.8 \textwidth]{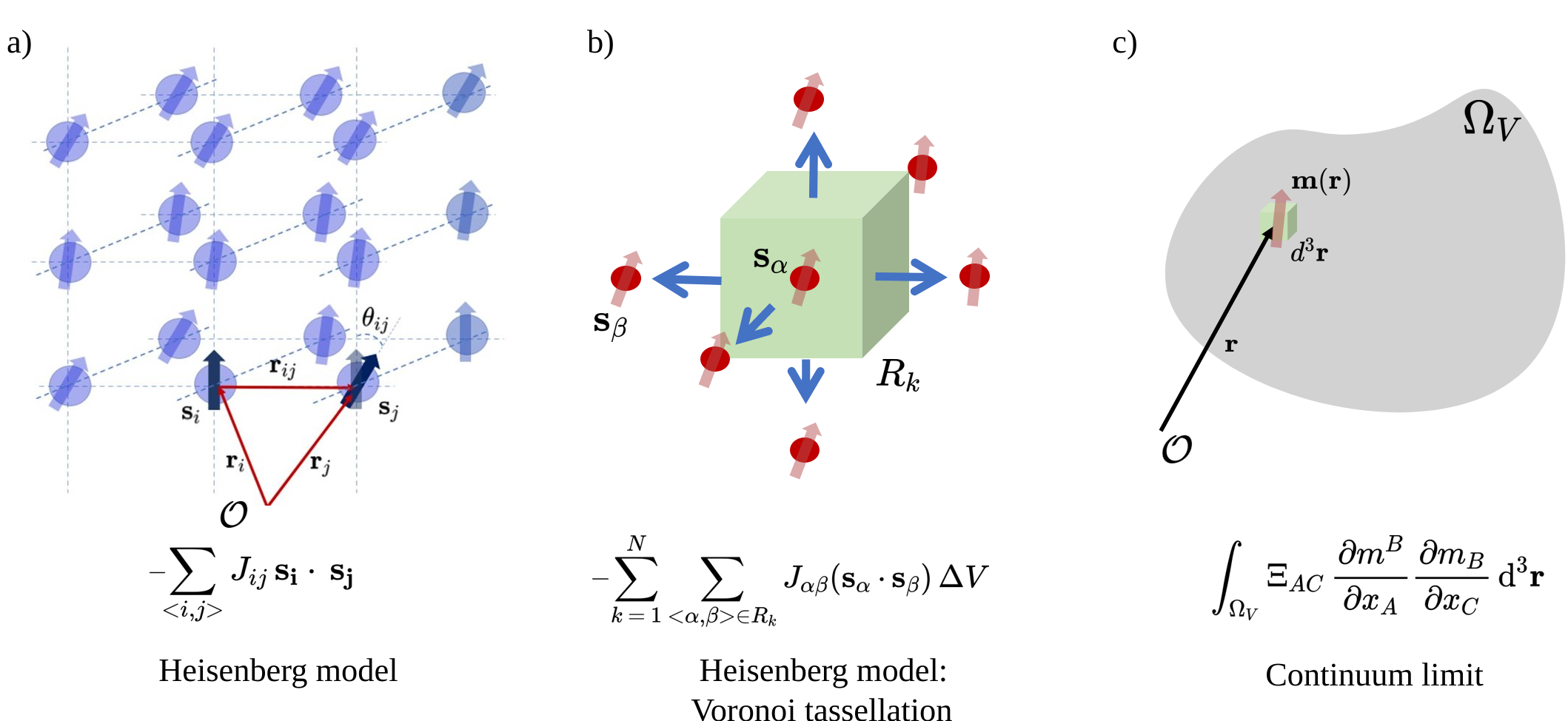}
    \caption{a) Heisenberg model on a cubic lattice.  We indicate the lattice sites with indices $i , j$ b) Decomposition of the cubic lattice in nearest neighbour clusters using the Voronoi tassellation \cite{GOO-17}. In this decomposed lattice we use the $k$ index to identify the cell and $\alpha , \beta$ indices to indicate the nearest neighbors. c) Continuum limit performed on the cell from b). The cell index $k$ becomes continuous $\upd^3 \bm{r}$ and the atomic moments become a continuous function of space $\bm{m}(\bm{r})$ }
    \label{fig:WS_Cell}
\end{figure*}

We start by writing the Heisenberg exchange interaction in the usual way
\begin{eqnarray}
\mathcal{H} = -\sum_{<i,j>} J_{ij} \bm{s}_i \cdot \bm{s}_j 
\end{eqnarray}
here $J\left(\bm{r}_{i j}\right)=J_{i j}$  are the coupling coefficients coming from the exchange integral, a function of the distance $\bm{r}_{i j}$ from atom $i$ to atom $j$.
We can split the energy sum in the contribution coming from each individual Wigner-Seitz (WS) cell of the lattice labelled $R_k$ (see Fig.\ref{fig:WS_Cell}-b). The advantages of this decomposition are twofold: firstly, the WS cell construction is based in the Voronoi tassellation \cite{GOO-17}, i.e. on the location of nearest neighbours. Secondly, the WS cell (a primitive cell) for a given lattice point inherits the full point group symmetry of the lattice by constructions \cite{Ashcroft76}
\begin{equation}
\mathcal{H} = -\sum_{k = 1}^{N} \sum_{<\alpha,\beta> \in R_k} J_{\alpha \beta} \, \bm{s}_\alpha \cdot \bm{s}_\beta \, \Delta V
\end{equation}
here $R_k$ represents the set of nearest neighbors used for the construction of the $k$-th WS cell \cite{PhysRev.43.804,Ashcroft76}. We define $\mathcal{E}_k$ as the energy density cost due to spin misalignment per WS cell with volume $\Delta V$ 
\begin{equation}
    \mathcal{E}_k = -\sum_{<\alpha,\beta>  \in R_k} J_{\alpha \beta} \, \bm{s}_\alpha \cdot \bm{s}_\beta
\end{equation}
we assume that $\bm{s}_{\alpha}$ and $\bm{s}_{\beta}$ are nearly parallel when nearest neighbours and have fixed length $\vert\bm{s}\vert^2=1$ (see Fig.\ref{fig:WS_Cell}-b), allowing us to write the exchange term as
\begin{eqnarray}
\mathcal{E}_k &=&   -\sum_{<\alpha,\beta>  \in R_k} J_{\alpha \beta}  \vert \bm{s} \vert^2 \cos \theta_{\alpha\beta} \,\,,\,\, \theta_{\alpha \beta} \ll 1 \\
&\approx& -\sum_{<\alpha,\beta>  \in R_k}J_{\alpha \beta}\delta^{\alpha \beta}-J_{\alpha \beta} \vert \bm{s}_{\alpha}-\bm{s}_{\beta}\vert^2.  \label{e}
\end{eqnarray}
Since we are treating the ferromagnetic case (i.e $J_{\alpha \beta}$ symmetric positive semi-definite) we can rewrite the exchange matrix as $J_{\alpha \beta} = G_{\alpha n}G_{n \beta} = \hat{\bm{G}}^T\hat{\bm{G}}$. We neglect the first term of \eqref{e} as it's simply a constant and we can rewrite the exchange energy as
\begin{equation}
    \mathcal{E}_k =  \sum_{<\alpha,\beta>  \in R_k} G_{\alpha n} G_{n \beta} \vert \, \bm{s}_{\alpha}-\bm{s}_{\beta}\vert^2 \label{exchange_dens}
\end{equation}
since for small $\theta_{i j} \approx \vert\bm{s}_{i}-\bm{s}_{j}\vert$, we assume that $\bm{s}_{i}$ can be fitted to a continuous function, i.e. $\bm{s}_i \rightarrow \bm{m}(\bm{r} - \bm{r}_i)$ of position in the lattice and that, to a sufficient approximation (see eq.(\ref{discr_Taylor}) in Appendix A),
\begin{equation}
\bm{m}(\bm{r} + \bm{r}_i  ) \approx \bm{m}(\bm{r}) + C_{i \alpha}\bm{l}^{\alpha} \cdot \bm{\nabla}\bm{m}(\bm{r}) \label{Adjacency}
\end{equation}
where $C_{i \alpha}$ represents the incidence matrix of the directed graph describing the nearest neighbor cluster of magnetic atoms (see the Supplementary Material for the case of a simple cubic lattice), $\bm{l}^\alpha=l^\alpha_A\hat{\bm{e}}^A$ represents the edge vectors of the directed graph and $\bm{\nabla} = \hat{\bm{e}}^A \frac{\partial}{\partial x^A}$.
With this generalized notation we can write the exchange term of eq.\eqref{e} as
\begin{equation}
\label{eq:semicontinuous_limit_discrte_energy}
\mathcal{E}(\bm{r})  =  \! \! \sum_{<\alpha,\beta>} \! \! \big[G_{\alpha n}C_{\gamma}^{\alpha}l^\gamma_A\frac{\partial m^B}{\partial x^A}\big]
\big[G_{n \beta}C_{\beta}^{\rho}l^\rho_C\frac{\partial m_B}{\partial x^C}\big]. 
\end{equation}
We remark how by transforming the magnetization in a continuous function of space, we automatically promote the energy per cell to a continuous function of space as well, i.e. $\mathcal{E}_k \rightarrow \mathcal{E}(\bm{r})$. This also applies to the volume $R_k$ of eq.\eqref{exchange_dens} which is now promoted to an infinitesimal volume element $d^3\bm{r}$. We store all the information related to the symmetry of the lattice and the exchange $J_{ij}$ in the (symmetric) anisotropic exchange tensor $\Xi_{AC}$.
\begin{equation}
    \Xi_{AC} := \sum_{<\alpha,\beta>} G_{\alpha n}C_{\gamma}^{\alpha}l^\gamma_A G_{n \beta}C_{\beta}^{\rho}l^\rho_C .  \label{eq:metric_tilde}
\end{equation}
The total exchange energy is now obtained by integrating $\mathcal{E}(\bm{r})$ over the whole volume
\begin{equation}
    E_{ex}[\bm{m}, \nabla \bm{m}] = \displaystyle{ \int_{\Omega_V} \Xi_{AC}  \,\frac{\partial m_B}{\partial x^C}\frac{\partial m^B}{\partial x^A}} \, \upd ^3\bm{r}. \label{eq:exch_comp}
\end{equation}
In the case of a cubic lattice all the exchange integrals $J_{ij} = J$ are constant and we have $\Xi_{AC} = -2 J \delta_{AC}$. We refer to the Supplementary material for the detailed calculation of the simple cubic case and the $C_{6v}$ case.
\begin{figure}
    \centering
    \includegraphics[scale=0.15]{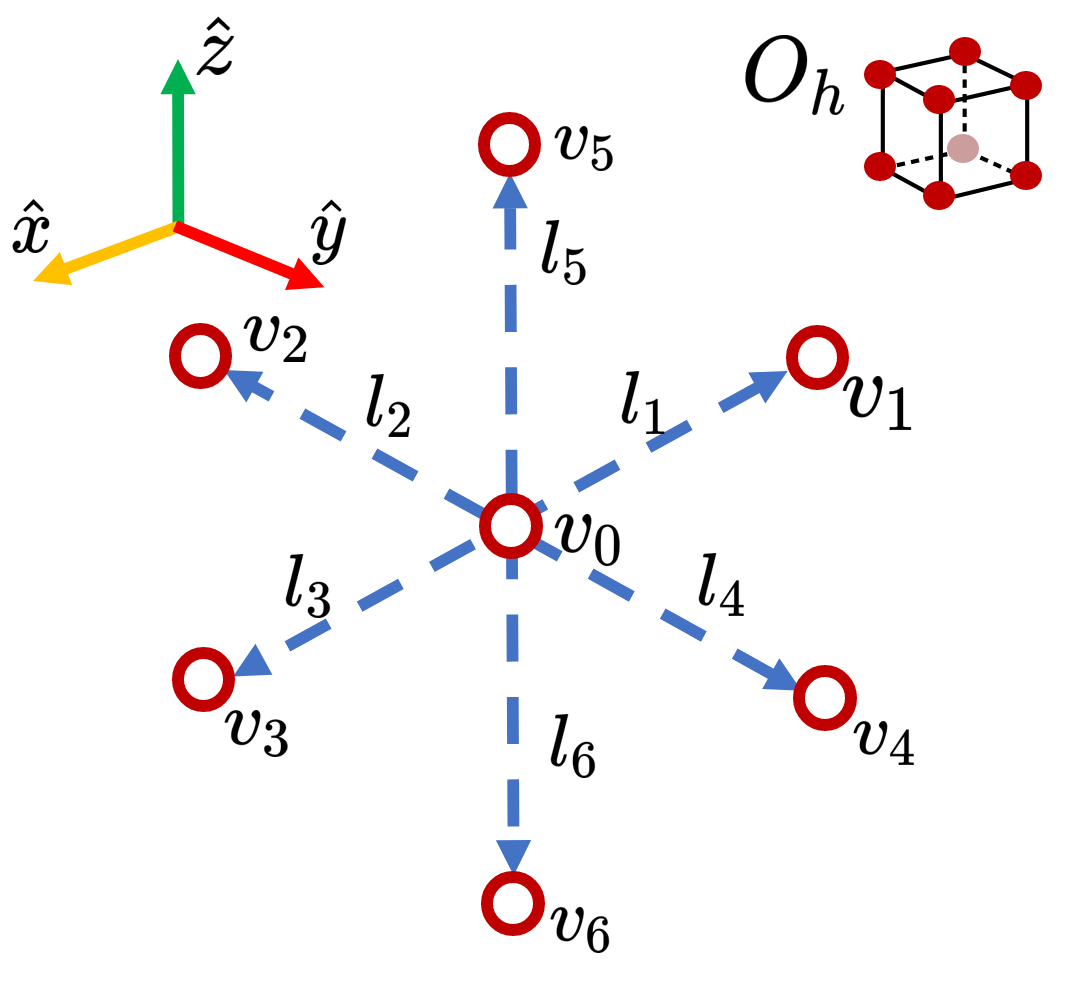}
    \caption{Directed graph representing the nearest neighbors (n.n) on the cubic lattice. The vertices are represented by $v_i$ and the edges by $l_i$.}
    \label{fig:cube}
\end{figure}
\section{Gauge covariant derivatives and the DMI tensor \label{Sec:Gauge_deriv}}
As discussed in \cite{Ansalone2021,Ansalone2022,Guslienko2016}, the appearance of DMI in a continuum theory of magnetic interactions is a direct consequence of promoting the global $SO(3)$ symmetry of the micromagnetic energy functional to a local symmetry.  The requirement of invariance with respect to a local rotation of the magnetic moment requires the inclusion of a non-Abelian Gauge degree of freedom encoded in the modification of the ordinary differential operator $\partial^i := \frac{\partial}{\partial x^i}$. Formally speaking, let $\mathcal{R}(\bm{x})$  be an element of SO(3) acting on 3-component vectors such as the local magnetization according to $\bm{m}' = \mathcal{R}(\bm{x})\bm{m}$. Enforcing invariance of the exchange energy density of  eq.\eqref{eq:exch_comp} requires us to redefine the differential operator via the covariant derivative ${\mathcal{D}}$ and the k-th component of the non-Abelian gauge potential $\bm{\mathcal{A}}_k={\mathcal{A}}^{Al}\varepsilon_{lk}^B\hat{\bm{e}}_A\otimes\hat{\bm{e}}_B$, in the following way \cite{Ansalone2021}:
\begin{equation}
    \partial^A m^B \rightarrow {\mathcal{D}}^A m^B = \partial^A m^B -  {\mathcal{A}}^{Al}\varepsilon_{lk}^{\phantom{ij}B}m^k,
\end{equation}
the non-Abelian gauge potential $\bm{\mathcal{A}}_k$ transforms according to the following rule
\begin{equation}
\bm{\mathcal{A}}'_k  = \mathcal{R}^T \bm{\mathcal{A}}_k\mathcal{R}+ \mathcal{R}^T \partial_k\mathcal{R}
\end{equation}
where the rotation matrices $\mathcal{R}$ can be represented via 
\begin{equation}
    \mathcal{R}(\bm{x}) = \exp(i \bm{\psi}(\bm{x}) \cdot \bm{J}),
\end{equation}
with $\bm{\psi}(\bm{x})=\psi^l(\bm{x})\hat{\bm{e}}_l$ and the $\psi^l(\bm{x})$ quantifies the amount of the local rotation in the direction $\hat{\bm{e}}_l$, and the generators of SO(3) are encoded in a vector $\bm{J} \text{ such that } [J_\rho , J_\sigma] = i \varepsilon_{\rho \sigma}^{\,\,\,\,\, \nu}J_\nu$. If we limit ourselves to the pure gauge case \cite{Ansalone2022}, we can restrict the gauge transformations to 
\begin{equation}
\bm{\mathcal{A}}'_k  = \mathcal{R}^T \partial_k\mathcal{R}
\end{equation}
And obtain the gauge covariant derivative of the form:
\begin{equation}
\mathcal{D}^A m^B = \partial^A m^B-\partial^A \psi^l \, \varepsilon_{lk}^{\phantom{ij}B}m^k
\end{equation}
Inserting this definition in eq.\eqref{eq:semicontinuous_limit_discrte_energy} yields the following expression for the energy density
\begin{equation}
\mathcal{E}(\bm{r}) = \!\!\sum_{\small{<\alpha,\beta>}}\! \!\big[G_{\alpha n}C_{\gamma}^{\alpha}l^\gamma_A\mathcal{D}^A m^B\big]
\big[G_{n \beta}C_{\beta}^{\rho}l^\rho_C\mathcal{D}^C m_B\big]
\end{equation}
and the following expression for the integrated total energy
\begin{align}
    &E_{ex}[\bm{m} ,\nabla \bm{m}] = \displaystyle \int_{\Omega_V} \upd^3 \bm{r} \, \{ \Xi_{AC}  \,\mathcal{D}^A m^B  \mathcal{D}^C m_B \} \\
     &= \displaystyle \int_{\Omega_V} \upd^3 \bm{r} \big\{ \, \,  \Xi_{AC}  \,\partial^A m^B\partial^C m_B - \nonumber \\ &- \Xi_{AC}  \, \partial^C \psi^l \varepsilon_{lkB}  \underbrace{\big[ (\partial^A m^B)m^k - m^B (\partial^A m^k) \big]}_{\mathcal{L}^{ABk}} \nonumber \\
   &+ \Xi_{AC} \,\,\partial^A \psi^l m^k \varepsilon_{lkB} \partial^C \psi^r m^s \varepsilon_{rsB} \big\} \label{exchange_expanded}
\end{align}
where $\Xi_{AC}$ is defined in eq.\eqref{eq:metric_tilde}.
We highlight how $\mathcal{L}_{A}^{\,\,\,Bk}$ represents the usual Lifshitz invariant terms of DMI. If we concentrate on terms of order $\mathcal{O}(\nabla \bm{\psi})$ we can rewrite the exchange in eq.\eqref{eq:exch_comp} as
\begin{align}
    E_{ex} &= \displaystyle \int_{\Omega_V} \upd^3 \bm{r} \big\{ \,  \Xi_{AC}  \,\partial^A m^B \partial^C m_B -   \Gamma^A_{\,\,\,kB}\mathcal{L}_{A}^{\,\,\,Bk} \, \big\} \\  &+ \mathcal{O}((\nabla \bm{\psi})^2)
\end{align}
where $\Xi_{AC}$ is defined in eq.\eqref{eq:metric_tilde}.
We highlight how $\mathcal{L}^{ABk}$ represents the usual Lifshitz invariant terms of DMI. If we concentrate on terms of order $\mathcal{O}(\nabla \bm{\psi})$ we can rewrite the exchange in eq.\eqref{eq:exch_comp} as
\begin{align}
    E_{ex} &= \displaystyle \int_{\Omega_V} \upd^3 \bm{r} \big\{ \,  \Xi_{AC}  \,\partial^A m^B \partial^C m_B -   \Gamma^A_{\,\,\,kB}\mathcal{L}_{A}^{\,\,\,Bk} \, \big\} \\  &+ \mathcal{O}((\nabla \bm{\psi})^2)
\end{align}
where we have introduced the compact notation 
\begin{equation}
    \Gamma_{AB}^k := \Xi_{AC} \, \partial^A \psi^l \varepsilon_{\,\,lB}^k. \label{eq:Gamma_pre}
\end{equation}
We can now apply the Neumann principle of crystallography \cite{Bhagavantam1967} to the $ \Gamma^A_{\,\,kB}$ prefactors of the Lifshitz invariants in eq.\eqref{eq:Gamma_pre} to reveal their independent components. Let $\mathcal{R}^{(\alpha)}$ be the 3-dimensional representation of the point group symmetry associated with the crystal system we are considering ($\alpha$ represents the index numbering the generators of the group). The Neumann principle \cite{Bhagavantam1967} imposes
\begin{eqnarray}
    \Gamma^i_{\,\,j k} = \big(\mathcal{R}^{(\alpha)}\big)_{i^\prime}^i\big(\mathcal{R}^{(\alpha)}\big)_j^{j^\prime} \big(\mathcal{R}^{(\alpha)}\big)_k^{k^\prime} \Gamma^{i^\prime}_{\,\,j^\prime k^\prime} \,\,,\,\, \forall \,\, \alpha . \label{Neumann}
\end{eqnarray}
We can now extract the non-vanishing components of the DMI tensor $Q_{Al} := \Xi_{AC} \, \partial^C \psi_l$ by contracting $\Gamma^k_{\,\,AB} $ with the Levi-civita  tensor $\varepsilon_{l}^{\,\,Bk}$ and using standard identity $ \varepsilon_{ijk}\varepsilon_{ljk} = 2 \delta_{il}$ which yields
\begin{equation}
    Q_{Al} = -\frac{1}{2} \Gamma^k_{\,\,AB}\varepsilon_{lk}^{\,\,\,\,B}. \label{DMI_tens}
\end{equation}
\begin{figure*}[h!]
    \centering
    \includegraphics[width = 0.8 \textwidth]{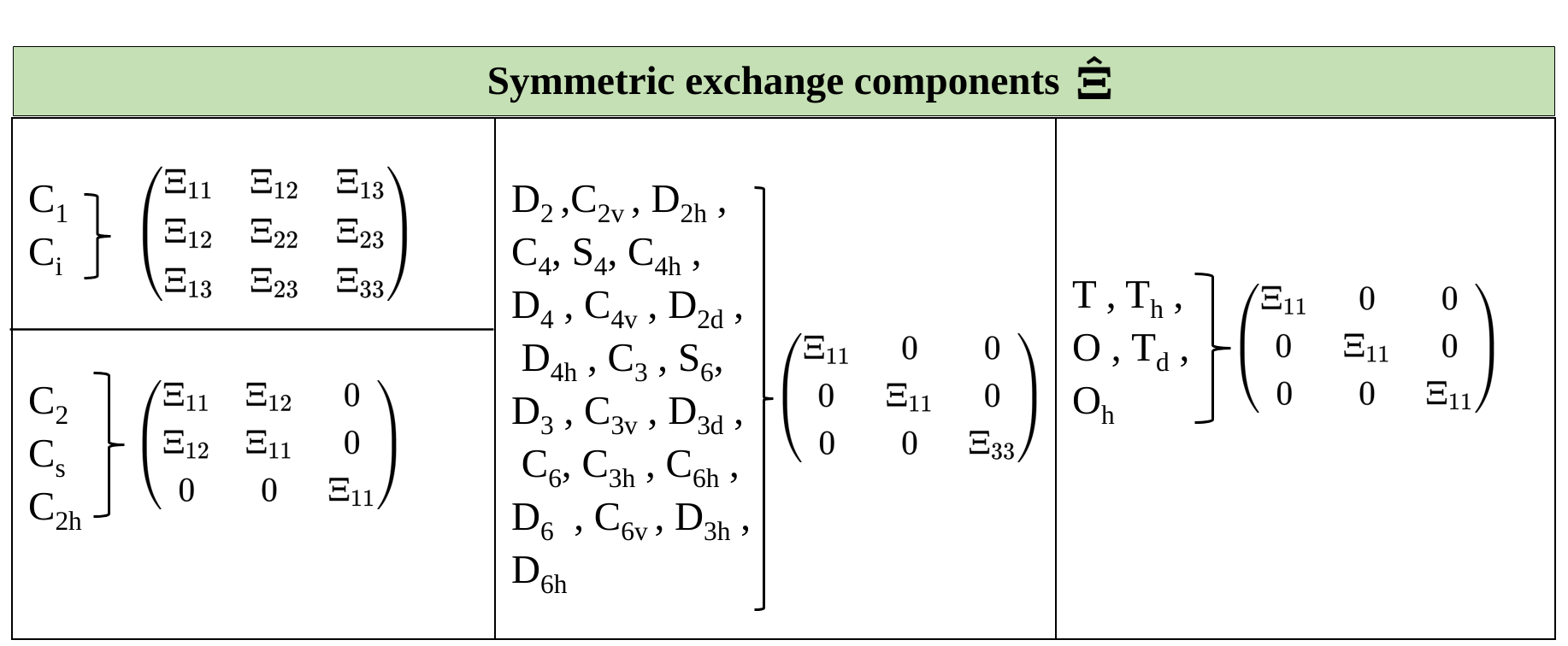}
    \caption{Symmetric exchange components $\hat{\bm{\Xi}}_{AC} := \Xi_{AC} $ from eq.\eqref{eq:metric_tilde} as a function of all 32 non-centrosymmetric crystallographic point groups as imposed by the Neumann principle \eqref{Neumann}. The generators are expressed in the basis used in \cite{Birss1966}.}
    \label{tab:g_tensor}
\end{figure*}
\begin{figure*}[h!]
    \centering
    \includegraphics[width = 0.8 \textwidth]{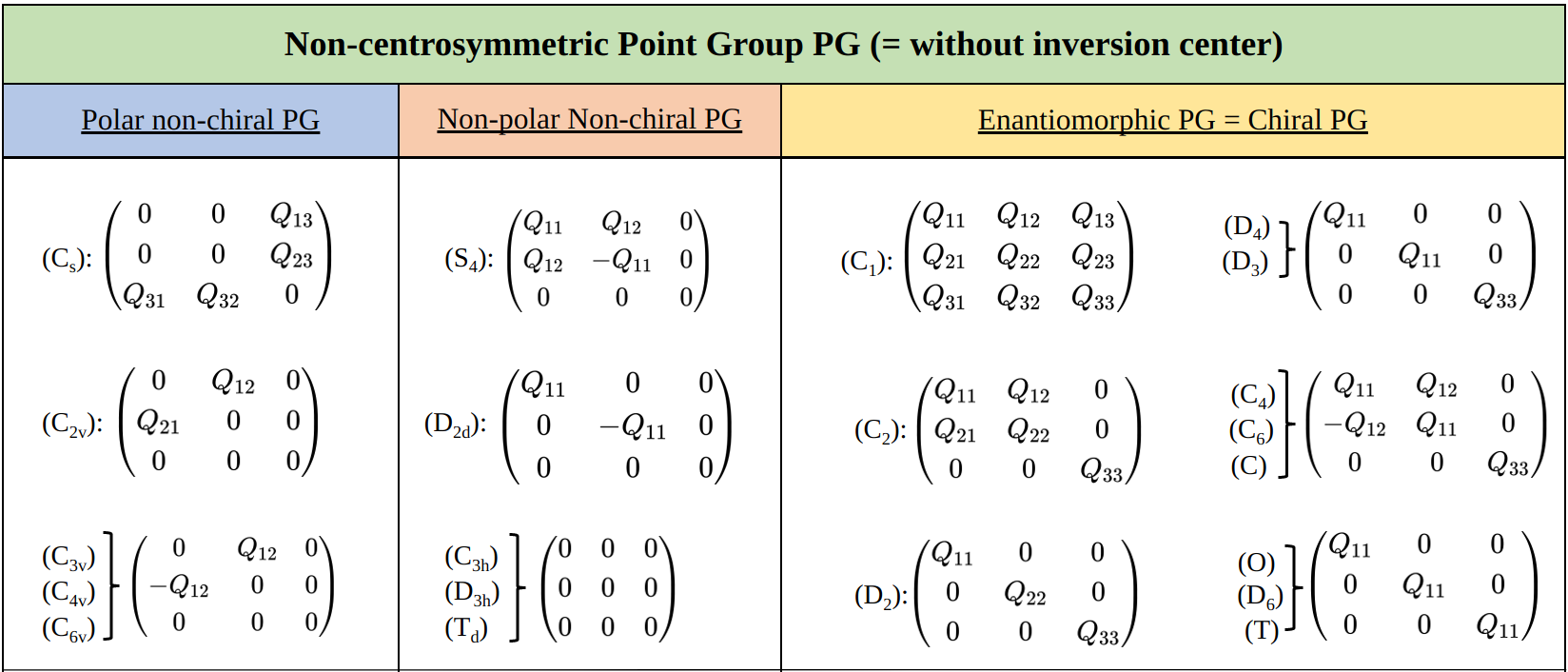}
    \caption{DMI tensor components $(\hat{\bm{Q}})_{Al} := Q_{Al}$ from eq.\eqref{DMI_tens} as a function of all 21 non-centrosymmetric crystallographic point groups as imposed by the Neumann principle of eq.\eqref{Neumann}. The 11 centrosymmetric point groups have a vanishing DMI tensor and are not shown. The generators are expressed in the basis used in \cite{Birss1966}.}
    \label{tab:DMI_tensor}
\end{figure*}
This formula now allows us to systematically predict the shape of the DMI tensor for all 32 crystallographic point groups. Using the generators for the crystallographic point groups contained in \cite{Aroyo2006} we arrive at DMI tensors $Q_{Al}$ of the form shown in Fig.\ref{tab:DMI_tensor}. With this we get the final form of the exchange energy functional of eq.\eqref{eq:main_res}
\begin{equation}
    E_{ex} = \displaystyle \int_{\Omega_V}  \big\{ \, \Xi_{AC}  \,\partial^A m^B  \partial^C m_B -  Q_{Al}\mathcal{M}^{lA} \, \big\} \, \upd^3 \bm{r} 
\end{equation}
where the chirality is given by $\mathcal{M}^{lA}  = \varepsilon^l_{\,\,kB}\mathcal{L}^{ABk}$. We remark how the Neumann principle can also be applied to the $\hat{\bm{\Xi}}$ tensor from eq.\eqref{eq:metric_tilde} 
\begin{equation}
    \Xi_{ij} =  \big(\mathcal{R}^{(\alpha)}\big)_i^{i^\prime} \big(\mathcal{R}^{(\alpha)}\big)^{j^\prime}_j \Xi_{i^\prime j^\prime}
\end{equation}
revealing the non vanishing components of symmetric exchange shown in Fig.\ref{tab:g_tensor}
\section{DMI tensor decomposition and ground state selection criterion}
We now proceed to describe some of the physical consequences of the symmetry properties of the DMI tensor written in the form of eq.\eqref{DMI_tens} \cite{Ansalone2022}. First of all we note that the DMI tensor of eq.\eqref{DMI_tens}, much like any rank-2 tensor, can be decomposed as a sum of symmetric and skew-symmetric components
\begin{equation}
\hat{\bm{Q}} = \underbrace{\frac{1}{2}(\hat{\bm{Q}} - \hat{\bm{Q}}^T )}_{\hat{\bm{Q}}_A} + \underbrace{\frac{1}{2}(\hat{\bm{Q}}^T + \hat{\bm{Q}} )}_{\hat{\bm{Q}}_S}. 
\end{equation}
A purely anti-symmetric DMI tensor yields Lifshitz invariant terms of the form 
\begin{equation}
    \mathcal{E}_{A;DMI} = -2 \bm{D}\cdot [ \bm{m}(\nabla \cdot \bm{m}) - (\nabla \cdot \bm{m})\bm{m}] \label{iDMI}
\end{equation}
where one expresses the anti-symmetric tensor as $(Q_A)_{ij} = D_k \varepsilon_{kij}$. This term corresponds to the continuum limit of the familiar microscopic interfacial DMI term \cite{Fert2013,PhysRevLett.115.267210}
\begin{equation}
    \mathcal{H}_{DMI} = \sum_{<i,j>} \bm{D}_{ij} \cdot (\bm{s}_i \times \bm{s}_j) 
\end{equation}
and is associated with chiral interactions coming from symmetry breaking along one direction.
Lifshitz invariant terms of the form eq.\eqref{iDMI} correspond to the surface DMI term appearing in magnetic thin films \cite{Thiaville2012}. The symmetric component of the DMI tensor yields an energy contribution of the form
\begin{equation}
    \mathcal{E}_{S;DMI} = - \bm{m}\cdot(\hat{\bm{Q}}_S\nabla \times \bm{m})
\end{equation}
Where $\hat{\bm{Q}}_S\nabla = (Q_S)_{ij}\partial_j$. The special case of a purely diagonal matrix yields an energy term of the form
\begin{equation}
    \mathcal{E}_{S;DMI} = - 2 (Q_S)_{ii}(\bm{m} \cdot \partial_i \bm{m})_i
\end{equation}
which, in the case of a single independent component $Q_{ii} = D\,\, \forall \,\,i$ yields 
\begin{equation}
    \mathcal{E}_{S;DMI} = - 2 D \, \bm{m} \cdot (\nabla \times \bm{m}). 
\end{equation}
This energy contribution corresponds to a bulk DMI term responsible for stabilizing bulk chiral structures \cite{Hoffmann,Nagaosa2013,Ansalone2022}. 
As discussed in \cite{Hoffmann}, the shape of the DMI tensor is a decisive factor in determining the appearance of skyrmionic or antiskyrmionic structures in the ground state of magnetic materials. The main result of \cite{Hoffmann} is the identification of the determinant of the DMI tensor as the relevant quantity predicting the stability of skyrmions or anti-skyrmions as follows:
\begin{equation}
    \det(\hat{\bm{Q}}) \begin{cases}
			< 0  & \text{Anti-skyrmions stabilized}\\
            > 0 & \text{Skyrmions stabilized} \\
            = 0 & \text{Coexistence}
		 \end{cases}. \label{eq:stability_crit}
\end{equation}
We can now apply this rule to the set of DMI tensors shown in Fig.\ref{tab:DMI_tensor} and discuss if the predicted ground state structure are compatible with experimental results discussed in the literature. 
\section{Discussion  \label{Sec:Discussion}}
The structure of eq.\eqref{DMI_tens} automatically implies that all centrosymmetric crystallographic point groups exclude the possibility to have DMI.
Furthermore, we notice how the above discussed method also correctly predicts the absence of DMI on some non-centrosymmetric crystal systems such as $T_d,C_{3h},D_{3h}$ which is in accordance with the literature \cite{Ullah2019}.
We can now proceed and discuss an example of a non vanishing DMI tensor: MnSi is a magnetic material which has attracted a lot of attention as it is one of the first materials in which the the presence of helical magnetic order was detected. As known from the literature, at low temperatures MnSi can be modelled using the extended classical Heisenberg model \cite{Hall2021} and the system is known to crystallize in a B20 structure \cite{Ishikawa1976}.
 If we isolate the magnetic atoms of this material, i.e. the Mn atoms, the resulting sublattice displays $T$ point group symmetry \cite{Rossler2006}. From Fig.\ref{tab:DMI_tensor}. we can see how the $T$ point group symmetry allows the material to have a purely diagonal DMI tensor which can be linked to the presence of bulk DMI, a form compatible with the appearance of bulk chiral magnetism \cite{Nagaosa2013,Ansalone2022,Ishikawa1976}. 
 \begin{eqnarray}
     (\hat{\bm{Q}}_T)_{ij}  &=& Q_{11} \delta_{ij} \\ &\Rightarrow& \mathcal{E}_{S;DMI} =- 2 Q_{11} \, \bm{m}\cdot(\nabla \times \bm{m}).
 \end{eqnarray}
If we want to consider only thin film geometries in which the growth direction lies parallel to the $\hat{z}$-axis, we simply have to consider the top left submatrix 
\begin{equation}
    \hat{\bm{Q}}_{T}[1,2;1,2] = \begin{pmatrix}Q_{11} & 0 \\0 &  Q_{11} \end{pmatrix}
\end{equation}
and the determinant of this submatrix is strictly positive. Recalling the selection criterion of eq.\eqref{eq:stability_crit} \cite{Hoffmann}, we know that a strictly positive determinant of the DMI tensor stabilizes skyrmions given the presence of a sufficiently high external Field \cite{Rossler2006}.
Further examples of applicability of the present formalism concern Heusler Alloys \cite{Galanakis2016,Singh2021,Elphick2021}. Despite the full Heusler structure displaying cubic symmetry (and therefore no DMI), recent studies have shown how altering the Mn concentration in inverse tetragonal Mn-based Heusler compounds such as Mn$_x$PtSn can lower the symmetry of the material as much as reaching D$_{2d}$ \cite{Swekis2021} in thin film geometries.  Observing the DMI tensor of the D$_{2d}$ symmetry group from Fig.\ref{tab:DMI_tensor}, we notice it has a symmetric traceless form: 
\begin{equation}
    \hat{\bm{Q}}_{D_{2d}} = \begin{pmatrix}0&Q_{12}&0\\Q_{12}&0&0\\0&0&0 \end{pmatrix}.
\end{equation}
Much like in the case of MnSi, restricting the DMI tensor to a thin film geometry in which the growth direction lies parallel to the $\hat{z}$-axis yields
\begin{equation}
    \hat{\bm{Q}}_{D_{2d}}[1,2;1,2] = \begin{pmatrix}0&Q_{12}\\Q_{12}&0 \end{pmatrix}.
\end{equation}
We immediately notice how $\det (\hat{\bm{Q}}_{D_{2d}}[1,2;1,2])= -Q_{12}^2 < 0 $. Again, according to eq.\eqref{eq:stability_crit} this DMI tensor can only stabilize antiskyrmions (given a sufficiently high external field). It has in fact been experimentally shown that Mn$_x$PtSn thin films can only support anti-skyrmions structures \cite{Singh2021}.
As a final case of interest, we consider heavy metal/ferromagnet (HM/FM) multilayers such as Pt(111)/Co which have been under intense scientific investigation for the development of energy efficient magnetic memory storage devices \cite{Yang2015,Deger2020}. In such systems, one of the effects of the inclusion/exclusion of the Pt layer is that of reducing the point group symmetry of the Co layer from $D_{6h}$ to $C_{3v}$  which, in terms of DMI tensors from Fig.\ref{tab:DMI_tensor} means
\begin{equation}
    \hat{\bm{Q}}_{D_{6h}} =\begin{pmatrix}0 & 0 & 0\\0 &  0 &  0\\0 &  0 & 0\end{pmatrix} \rightarrow \hat{\bm{Q}}_{C_{6v}} =  \begin{pmatrix}0 & Q_{12} & 0\\-Q_{12} &  0 &  0\\0 &  0 & 0\end{pmatrix}.
\end{equation}
This immediately highlights how the broken inversion symmetry can cause the emergence of DMI interaction at Pt/Co(111) interfaces. At the same time, we notice how $\det(\hat{\bm{Q}}_{C_{6v}}) > 0$ and therefore only skyrmions can be stabilized in systems displaying this symmetry \cite{Yang2015,wang2019construction}.
 As an aside, we also remark how the applications of strain gradients \cite{Kitchaev2018} and electric fields \cite{Basso2020a} to materials displaying chiral interactions can alter the properties of the DMI tensor. In particular, strong electric fields can lead to the change of the antisymmetric components of the DMI tensor \cite{Ansalone2021} and could therefore constitute a way to manipulate skyrmions and antiskyrmions in an energy efficient way. 
\section{Conclusions: }
In this work we presented a new perspective on the continuum limit of the classical Heisenberg model to derive the micromagnetic exchange energy functional. Our approach systematically keeps track of the crystal symmetries of the system and reveals the importance of the interplay between symmetric anisotropic exchange and DMI. We show a rigorous treatment of higher order interactions when promoting the symmetry of the Hamiltonian from global to local via the introduction of gauge covariant derivatives \cite{Ansalone2021,Ansalone2022}. As an example, we show how the symmetry constraints imposed by the lattice can be implemented rigorously via the Neumann principle of crystallography, revealing the independent components of the DMI tensor for all 32 crystallographic point groups. We point out how the determinant of the DMI tensor can be used as a tool to predict the stabilization of skyrmions and antiskyrmions \cite{Hoffmann} and we observe how several experimental results behave in accordance with our theoretical predictions.
\section{Appendix A: Taylor expansion with discrete calculus using graph theory }
In the following we review the Taylor expansion of discrete calculus \cite{Grady2010} in order to motivate our derivation of the continuum limit of micromagnetic exchange in the first section. Let us write down the general expression for the Taylor expansion of the magnetization vector field for small $\Delta \bm{r}$
\begin{equation}
    \bm{m}(\bm{r} + \Delta\bm{r}) \approx \bm{m}(\bm{r}) + d\bm{m}(\bm{r} ; \Delta\bm{r} ) + \mathcal{O}\big( \vert \Delta \bm{r}\vert^2 \big) \label{general_exp}
\end{equation}
where $ d\bm{m}(\bm{r} ; \Delta\bm{r} )$ represents the directional derivative of the vector valued function $\bm{m}(\bm{r}) $ along the vector $\Delta\bm{r}$, i.e.
\begin{equation}
    dm_i(\bm{r} ; \Delta\bm{r} ) = \nabla m_i \cdot \Delta\bm{r} = \partial_j m_i \Delta r_j.  \label{discr_Taylor}
\end{equation}
\noindent We now formally define a nearest-neighbour cluster of atoms as a directed graph - (see Fig.\ref{fig:cube} for an example of the cubic lattice) introducing the edge-node incidence matrix $C_{ij}$ \cite{Grady2010} defined as
\begin{equation*}
    C_{ij} = \begin{cases} 
    0 & \text{if edge } i \text{ and node } j \text{ are not connected} \\
    +1 & \text{if edge } i \text{ is directed toward node } j \,\, :  i \rightarrow j  \\
    -1 & \text{if edge } i \text{ is directed out of node } j  \,\, :  i \leftarrow j
    \end{cases}.
\end{equation*}
It can be shown \cite{Grady2010} that the incidence matrix of a graph is the natural matrix representation of the discrete differential. Let us now denote the edges of the directed graph representing the lattice as $\bm{l}^i$ (see Fig.\ref{fig:cube}), we can generalize eq. \eqref{discr_Taylor} to
\begin{equation}
    dm_i(\bm{r} ; \Delta\bm{r} ) = (\nabla_k m_i) C_{kj}\bm{l}^j 
\end{equation}
and finally write the components of the expansion \eqref{general_exp} as 
\begin{equation}
   m_i(\bm{r} + \Delta\bm{r}) \approx m_i(\bm{r}) + (\nabla_k m_i(\bm{r})) C_{kj}\bm{l}^j + \mathcal{O}(\nabla \bm{m}).
\end{equation}
\bibliographystyle{eplbib}

\section{Acknowledgement}
This project has received funding from the European Union’s Horizon 2020 research and innovation programme under the Marie Skłodowska-Curie grant agreement No. 860060 “Magnetism and the effect of Electric Field” (MagnEFi).

\end{document}


\title{Gauge theory applied to magnetic lattices - Supplementary material}

\shorttitle{Supplementary material} 

\author{A. Di Pietro\inst{1,2} \and P. Ansalone\inst{1} \and V. Basso \inst{1} \and A. Magni \inst{1} \and G. Durin \inst{1}}
\shortauthor{A. Di Pietro \etal}

\institute{                    
  \inst{1} INRIM - Strada delle Cacce 91, 10135 Torino, Italy\\
  \inst{2} Politecnico di Torino - Corso Duca degli Abruzzi, 24, 10129 Torino, Italy 
}

\maketitle

\section{Exact calculation of $\Xi_{AC}$ for the simple cubic lattice} \label{Appendix:A}
\begin{figure}
    \centering
    \includegraphics[scale=0.15]{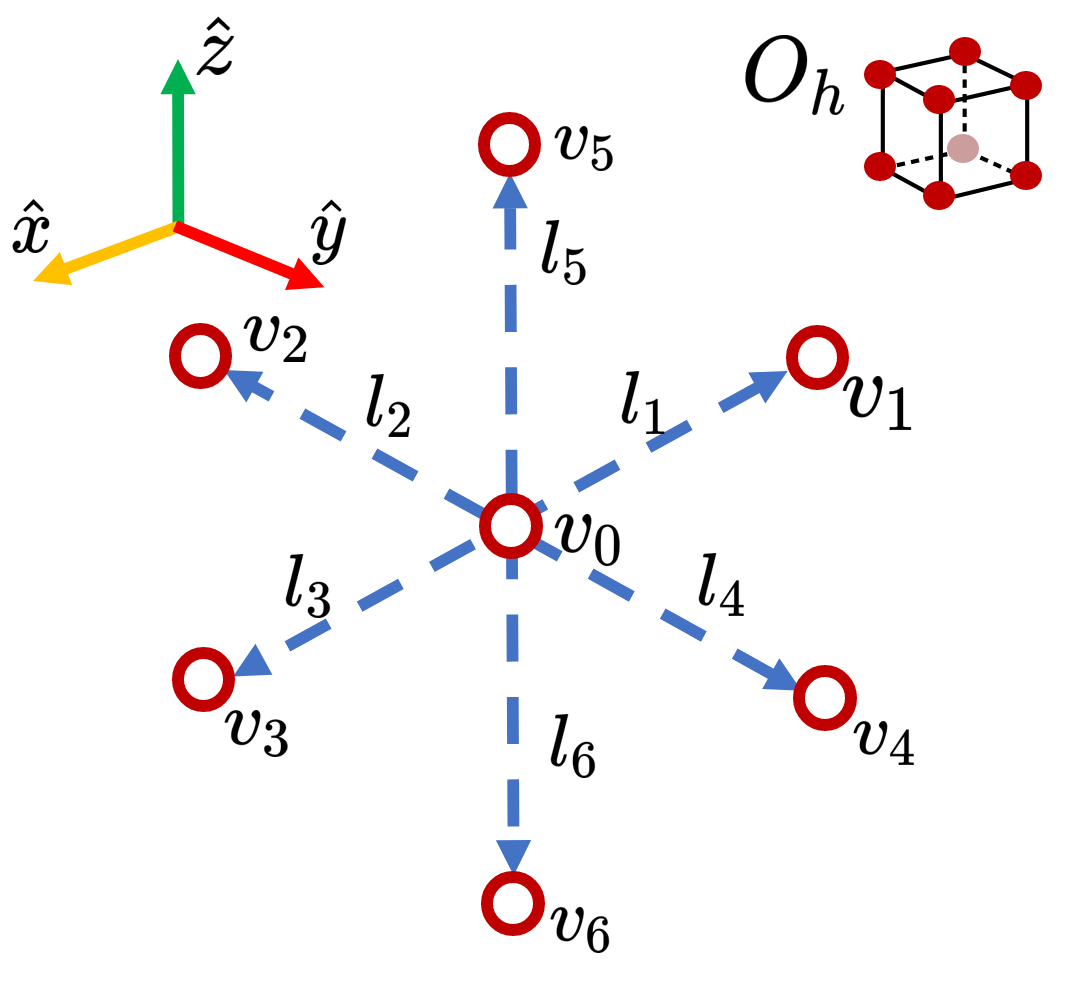}
    \caption{Directed graph representing the nearest neighbors (n.n) on the cubic lattice. The vertices are represented by $v_i$ and the edges by $l_i$.}
    \label{Supp:fig:cube}
\end{figure}
In the following we provide a step-by-step calculation of the $\Xi_{AC}$ tensor according to the definition
\begin{equation}
        \Xi_{AC} :=  \big(\underbrace{\sum_{i} G_{in}C_{\gamma}^{i}}_{\tilde{C}^n_\gamma}\big)(\bm{l}^\gamma)_A  \big(\underbrace{\sum_{j} G_{n j} C_{j}^{\rho}}_{\tilde{C}^\rho_n}\big)(\bm{l}^\rho)_C .
\end{equation}
As a first step we compute the incidence matrix of the oriented graph representing the nearest neighbors of the simple cubic lattice (Fig.\ref{Supp:fig:cube}). 
\begin{equation}
    C_{i j} := \hat{C} = \left(\begin{array}{ccccccc}
1 & -1 & 0 & 0 & 0 & 0 & 0 \\
1 & 0 & -1 & 0 & 0 & 0 & 0 \\
1 & 0 & 0 & -1 & 0 & 0 & 0 \\
1 & 0 & 0 & 0 & -1 & 0 & 0 \\
1 & 0 & 0 & 0 & 0 & -1 & 0 \\
1 & 0 & 0 & 0 & 0 & 0 & -1 \\
\end{array}\right).
\end{equation}
We can write the matrix denoting the edges of the graph  $ (\bm{l}^\gamma)^A$ as
\begin{equation}
    \hat{R} := (\bm{l}^\gamma)_A = \left(\begin{array}{ccc}1&0&0\\0&1&0\\-1&0&0\\0&-1&0\\0&0&1\\0&0&-1\\\end{array}\right)
\end{equation}
The $\tilde{C}$ matrix can be interpreted as a weighted directed graph, where the weight on each edge of the graph is determined by the magnitude of the exchange integral $J_{ij}$. If we assume that the strength of the interaction is a function of the inter-site distance, we have $J_{ij} = J$  and $\tilde{C}$ acquires the simplified form
\begin{equation}
    \hat{\tilde{C}} = \omega \cdot \hat{C} \, , \, \omega^2 = J .
\end{equation}
We can now write the the expression $\tilde{g}^{AC}$ in matrix form
\begin{eqnarray}
    \Xi_{AC}&:=&  \big(\underbrace{\sum_{i} G_{in}C_{\gamma}^{i}}_{\tilde{C}^n_\gamma}\big)(\bm{l}^\gamma)_A  \big(\underbrace{\sum_{j} G_{n j} C_{j}^{\rho}}_{\tilde{C}^\rho_n}\big)(\bm{l}^\rho)_C   \label{metric_tilde} \\ 
   &=& (\bm{l}_\rho)_C \tilde{C}^\rho_n \tilde{C}^n_\gamma (\bm{l}^\gamma)_A \\ &=&  J (\hat{R}^T \cdot \hat{C} \cdot \hat{C}^T \cdot \hat{R})
\end{eqnarray} 
with these definitions we can proceed and compute the $\Xi_{AC}$ tensor 
\begin{equation}
\Xi_{AC} = -2 J \delta_{AC}.
\end{equation}
\section{ Effect of the basis choice on the DMI tensor} \label{Appendix:C}
\begin{figure*}[h!]
    \centering
    \includegraphics[scale=0.185]{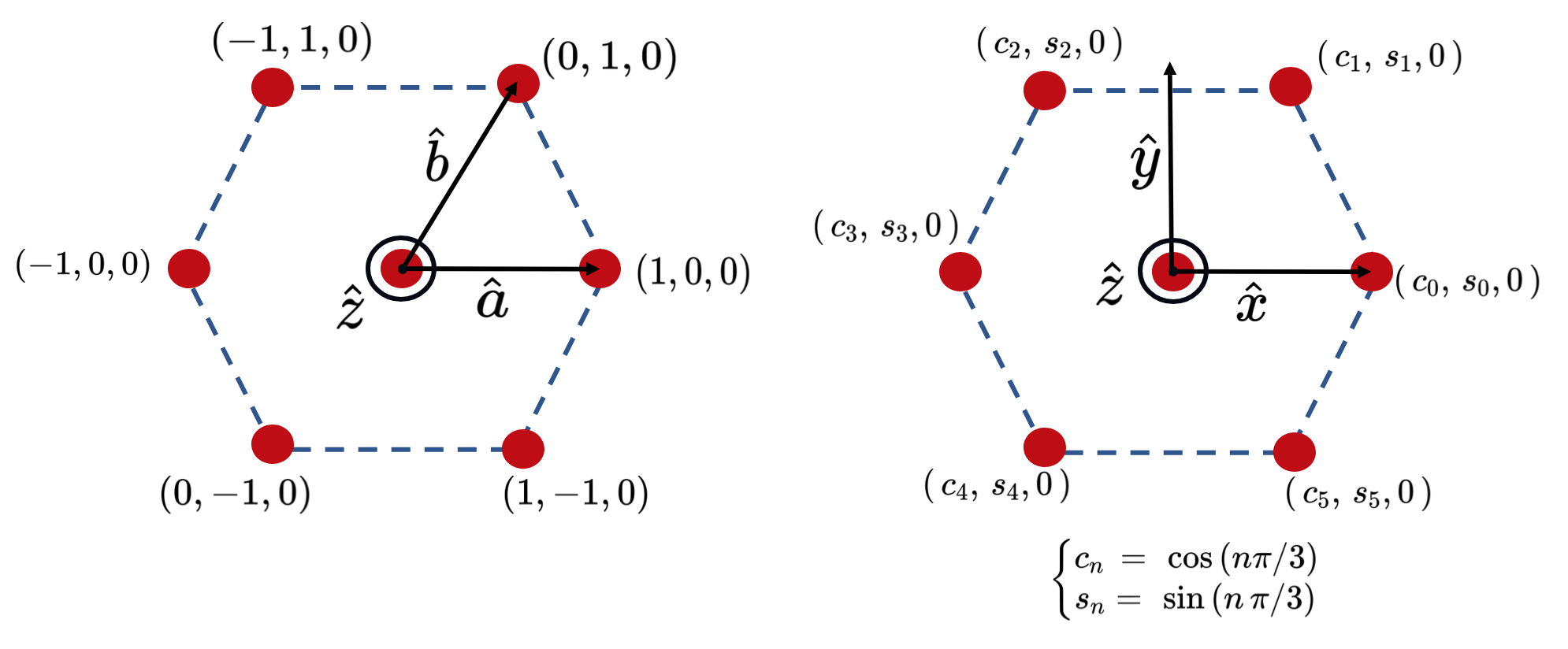}
    \caption{2 different basis choices to represent the lattice vectors in the $C_{6v}$ geometry. We only depict the basis choice in the $C_6$ plane as along the $z$-axis the choice is trivial.}
    \label{fig:basis_choice}
\end{figure*}

The choice of the basis can have an effect on the form of the DMI tensor and should therefore be treated carefully. As an example we show the case of $C_{6v}$ symmetry.
\begin{figure}
    \centering
    \includegraphics[scale=0.15]{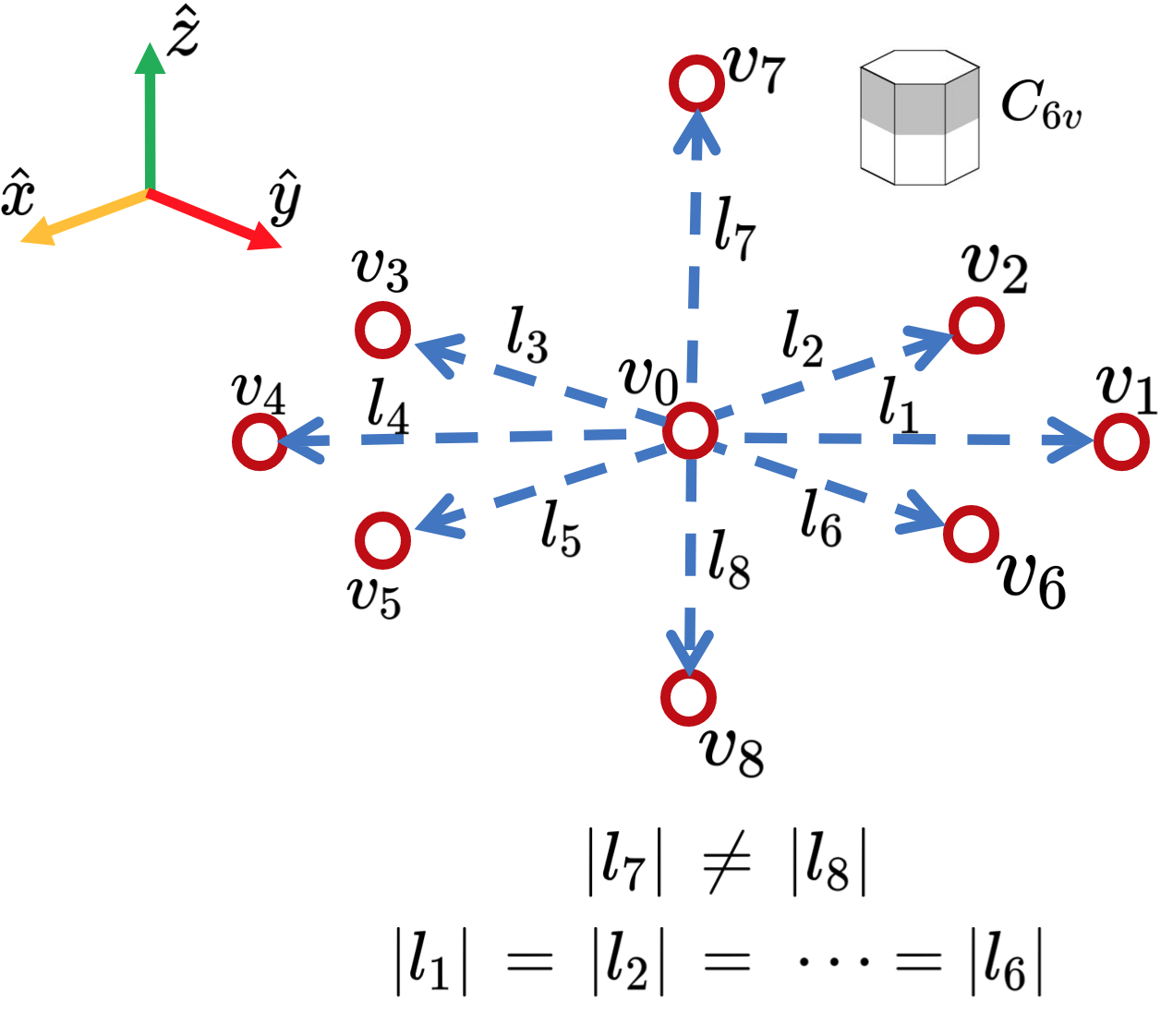}
    \caption{Directed graph representing the nearest neighbors (n.n) on a $C_{6v}$. The vertices are represented by $v_i$ and the edges by $l_i$.}
    \label{fig:C6v}
\end{figure}
The edge-node incidence matrix is given by
\begin{equation}
\begin{scriptsize}
\hat{C} = \left(\begin{array}{ccccccccc}1&-1&0&0&0&0&0&0&0\\1&0&-1&0&0&0&0&0&0\\1&0&0&-1&0&0&0&0&0\\1&0&0&0&-1&0&0&0&0\\1&0&0&0&0&-1&0&0&0\\1&0&0&0&0&0&-1&0&0\\1&0&0&0&0&0&0&-1&0\\1&0&0&0&0&0&0&0&-1\\\end{array}\right).
\end{scriptsize}
\end{equation}
At this point we can write down the crystal vectors in 2 different bases depicted in Fig.\ref{fig:basis_choice} 
\begin{equation}
\begin{scriptsize}
    \hat{R}_{\hat{a},\hat{b},\hat{z}} = \left(\begin{array}{ccc}1&0&0\\0&1&0\\-1&1&0\\-1&0&0\\0&-1&0\\1&-1&0\\0&0&c\\0&0&-\tilde{c}\\\end{array}\right) \,\,\, \hat{R}_{\hat{x},\hat{y},\hat{z}} = \left(\begin{array}{ccc}1&0&0\\\frac{1}{2}&\frac{\sqrt{3}}{2}&0\\-\frac{1}{2}&\frac{\sqrt{3}}{2}&0\\-1&0&0\\-\frac{1}{2}&-\frac{\sqrt{3}}{2}&0\\\frac{1}{2}&-\frac{\sqrt{3}}{2}&0\\0&0&c\\0&0&-\tilde{c}\\\end{array}\right) 
\end{scriptsize}
\end{equation}
where $c$ and $\tilde{c}$ represent the distance of nearest neighbors in the $\pm z$ direction. The different choice of basis yields different $\Xi_{AC}$ tensors
\begin{eqnarray}
    \hat{\bm{\Xi}}_{\hat{a},\hat{b},\hat{z}} &=& \left(\begin{array}{ccc}-4J&2J&0\\2J&-4J&0\\0&0&-2(c -\tilde{c})^2J\\\end{array}\right) \\ \hat{\bm{\Xi}}_{\hat{x},\hat{y},\hat{z}} &=&  \left(\begin{array}{ccc}-3J&0&0\\0&-3J&0\\0&0&-2(c -\tilde{c})^2J\\\end{array}\right).
\end{eqnarray}
This of course has implications for the shape of the DMI tensor
\begin{eqnarray}
    \hat{\bm{Q}}_{C_{6v},\hat{a},\hat{b},\hat{z}} &=& \begin{pmatrix}Q_{11} & 2Q_{11} & 0\\-2Q_{11} &  -Q_{11} &  0\\0 &  0 & 0\end{pmatrix} \\  \hat{\bm{Q}}_{C_{6v},\hat{x},\hat{y},\hat{z}} &=& \left(
\begin{array}{ccc}
 0 & Q_{12} & 0 \\
 -Q_{12} & 0 & 0 \\
 0 & 0 & 0 \\
\end{array}
\right) \label{eq:DMI_Tensor_Comp}.
\end{eqnarray}
This discrepancy can be reconciled by recalling the definition of the DMI tensor obtained in the main text of the article, namely 
\begin{equation}
    Q_{Al} = \Xi_{AC} \partial^C \psi_l 
\end{equation}
from which we have
\begin{equation}
    (\Xi)^{-1}_{AC} Q_{Al} =  \partial^C \psi_l.
\end{equation}
We notice how the two DMI tensor shapes can be connected one to another via 
\begin{align}
    \hat{\bm{\Xi}}^{-1}_{\hat{a},\hat{b},\hat{z}} \hat{\bm{Q}}_{C_{6v},\hat{a},\hat{b},\hat{z}} = \hat{\bm{\Xi}}^{-1}_{\hat{x},\hat{y},\hat{z}} \hat{\bm{Q}}_{C_{6v},\hat{x},\hat{y},\hat{z}}
\end{align}
To avoid any confusion, we always chose the reference frame and the generators reported in \cite{Birss1966}.

\bibliographystyle{eplbib}